\documentclass{svproc}
\usepackage{changes}
\usepackage{authblk} %for authers
\usepackage{url}

\usepackage{amsmath}

\usepackage{graphicx}
\usepackage{float}
\usepackage{subfig}
\usepackage{caption}
\definechangesauthor[name={Shahar}, color=red]{Shahar}

\newcommand{\remove}[1]{}

\begin{document}
	
	\mainmatter              % start of a contribution
	\title{ERC20 Transactions over Ethereum Blockchain: Network Analysis and Predictions}

	\author{Shahar Somin\inst{1,2}\and Goren Gordon\inst{2,3} \and Alex Pentland \inst{1} \and \\ Erez Shmueli \inst{2}\and Yaniv Altshuler\inst{1,3}
	}
	%\author{Yaniv Altshuler\vspace{-3.5mm}}
	%\affil{MIT Media Lab\\\texttt{yanival@mit.edu}}
	%
	%\author{Shahar Somin\vspace{-3.5mm}}
	%\affil{Endor.coin\\\texttt{shahar@endor.com}}
	%
	\authorrunning{S. Somin, G. Gordon, A. Pentland, E. Shmueli  and Y. Altshuler} % abbreviated author list (for running head)
	\tocauthor{S. Somin, G. Gordon, A .Pentland, E. Shmueli and Y. Altshuler}
	\institute{MIT Media Lab, Cambridge, MA, USA\\
		\email{$\textbf{shaharso} $@media.mit.edu}
		\texttt{}
		\and
		Industrial Engineering Department, Tel Aviv University, Israel\\
		%		\email{goren@gorengordon.com}
		\and
		Endor Ltd.}
	
	\maketitle              % typeset the title of the contribution
	
	\begin{abstract}
		
		Following the birth of Bitcoin and the introduction of the Ethereum ERC20 protocol a decade ago, recent years have witnessed a growing number of cryptographic tokens that are being introduced by researchers, private sector companies and NGOs. 
		The ubiquitous of such Blockchain based cryptocurrencies give birth to a new kind of rising economy, which presents great difficulties to modeling its dynamics using conventional semantic properties.
		Our work presents the analysis of the dynamical properties of the ERC20 protocol compliant crypto-coins’ trading data using a network theory prism.
		We examine the dynamics of ERC20 based networks over time by analyzing a meta-parameter of the network --- the power of its degree distribution. 
		Our analysis demonstrates that this parameter can be modeled as an under-damped harmonic oscillator over time, enabling a year forward of network parameters predictions.

	\end{abstract}
	\section{Introduction}
	Blockchain technology, which has been known by mostly small technological circles up until recently, is bursting throughout the globe, with a potential economic and social impact that could fundamentally alter traditional financial and social structures. 
	Launched in July 2015 \cite{buterin2014ethereum}, the Ethereum Blockchain is a public ledger that keeps publicly accessible records of all Ethereum related transactions. 
	The ability of the Ethereum Blockchain to store not only ownership, similarly to the Bitcoin Blockchain, but also execution code, in the form of \emph{"Smart Contracts''}, has recently led to the creation of an immense number of new types of "tokens'', based on the Ethereum ERC20 protocol. 

	Apart from providing full data of prices, volumes and holders distribution, the ERC20 transactional data also presents the monetary activity of anonymous individuals, which is otherwise scarce and hard to obtain due to confidentiality and privacy control.
	Thereby, this ERC20 digital ecosystem intrinsically provides a rare opportunity to analyze and model financial behavior in an evolving market over a long period of time.
	Specifically, understanding the governing forces upon this emerging economy, and in turn being able to perform accurate predictions of the economy's state are fundamental, as this market is becoming increasingly relevant to the traditional financial world. 
	
	In this work we aim to broaden our comprehension of the dynamics this financial ecosystem undergoes, from a network theory perspective. 
	Specifically, we first demonstrate how the dynamics of the degree distribution's power parameter can be modeled by an under-damped oscillator with zero-mean Gaussian noise.
	In turn, this analytical model enables us to predict network's $ \gamma $ dynamics, reliably predicting a whole year forward in time.

	\section{Background and Related Work}
	\label{sec.related}
	
	Blockchain's ability to process transactions in a trust-less environment, apart from trading its official cryptocurrency, the \emph{Ether}, presents the most prominent framework for the execution of "\emph{Smart Contracts}'' \cite{wood2014ethereum}. Smart Contracts are computer programs, formalizing digital agreements, automatically enforced to execute any predefined conditions using the consensus mechanism of the Blockchain, without relying on a trusted authority. They empower developers to create diverse applications in a Turing Complete Programming Language, executed on the decentralized Blockchain platform, enabling the execution of any contractual agreement and enforcing its performance.
	
	Moreover, Smart Contracts allow companies or entrepreneurs to create their own proprietary tokens on top of the Blockchain protocol \cite{catalini2018initial}. These tokens are often pre-mined and sold to the public through Initial Coin Offerings (ICO) in exchange of Ether, other crypto-currencies, or \emph{Fiat Money}. The issuance and auctioning of dedicated tokens assist the venture to crowd-fund their project's development, and in return, the ICO tokens grant contributors with a redeemable for products or services the issuer commits to supply thereafter, as well as the opportunity to gain from their possible value increase due to the project's success. The most widely used token standard is Ethereum’s \emph{ERC20} (representing Ethereum Request for Comment), issued in 2015. The protocol defines technical specifications giving developers the ability to program how new tokens will function within the Ethereum ecosystem.

	There has been a surge in recent years in the attempt to model social dynamics via statistical physics tools \cite{castellano2009statistical}, ranging from opinion dynamics, through crowd behaviors to language dynamics. The physical tools used are also varied, ranging from Ising models \cite{smug2018generalized} to topology analysis \cite{castellano2009statistical}.
	More specifically, previous studies have implemented physics-based approaches to the analysis of economic markets.
	Econophysics have attempted to describe the dynamical nature of the economy with different, and increasingly sophisticated physical models. Frisch \cite{frisch1933propagation}, who started this trend, has suggested to use a damped oscillator model to the economy post wars or disasters, with the assumption that there is an equilibrium state that has been perturbed. Since then, many new models have been suggested, ranging from quantum mechanical models \cite{ye2008non,gonccalves2013quantum} to chaos theory \cite{goodwin1993economy,puu2013attractors}.
	However, all of these models have attempted to describe the economy, represented by a singular {\em value}, e.g. stock market prices, whereas the underlying network of the economy has not been addressed.
	
	Network science, however, has exceedingly contributed to multiple and diverse scientific disciplines in the past two decades, by examining exactly diverse network related parameters. Applying network analysis and graph theory have assisted in revealing the structure and dynamics of complex systems by representing them as networks, including social networks \cite{barrat2008dynamical,newman2003structure,newman2005power}, computer communication networks  \cite{pastor2007int}, biological systems \cite{barabasi2004bio}, transportation \cite{shmueli2015ride,altshuler2015rationality}, IOT \cite{SBP-Crowd}, emergency detection \cite{altshuler2013social} and financial trading systems \cite{SBP-Trends,pan2012decoding,shmueli2014temporal}. 
	
	Most of the research conducted in the Blockchain world, was concentrated in Bitcoin, spreading from theoretical foundations \cite{bonneau2015sok}, security and fraud \cite{meiklejohn2013fistful,Pentland-Shreir-Shrobe2018} to some comprehensive research in network analysis \cite{ron2013quantitative,maesa2016uncovering,Lischke2016}. The world of Smart contracts has recently inspired research in aspects of design patterns, applications and security \cite{bartoletti2017empirical,anderson2016new,christidis2016blockchains,atzei2017survey}, policy towards ICOs has also been studied \cite{catalini2018initial}. Some preliminary results examining network theory's applicability to ERC20 tokens has been made in \cite{somin2018network,somin2018social}, specifically by validating that this financial ecosystem, when considered as a network of interactions, adheres to key network theory principles, such as power-law degree distribution.
	
	 In this paper we aim to examine how this prominent field can enhance the understanding of the underlying structure of the ERC20 tokens trading data, model it's stabilization process as a network over time and achieve predictive abilities.

	\section{Methodology}
	\label{sec.methodology}
	
	\subsection{Power-Law Fit}
	
	The degree distribution of a given graph is plotted on a double logarithmic scale, over 20 logarithmically spaced bins, between the minimal and maximal degrees of relevant graph. We've selected splitting the data along 20 bins, in order to accomodate both small networks, having small sets of vertices and consequently possibly small degree sequences, and also large networks obtaining much larger variance of the degree set.
	
	Several approaches are known in literature for fitting the power law distribution to a linear model in the double logarithmic scale and for estimating its \textit{goodness-of-fit}, see for example \cite{clauset2009power}. 
	In this paper, we have chosen to fit the bins' heights to a Linear Model, using ordinary Least Squares Regression, while considering all binned data points, and not only their tail. 
	We further chose to verify the \textit{goodness-of-fit} of the power-law model to the degree distribution by calculating the coefficient of determination of the fit, i.e its $ R^2 $, computed as follows:
	\begin{equation}
	R^2 = 1-\frac{\sum\limits_{k} (y_k-f_k)^2}{\sum\limits_{k}(y_k-\bar{y})^2}
	\end{equation}
	where $ y_k = P(k)$ are the degree distribution values, $ f_k $ are the modeled degrees by the fitted power-law model, and $ \bar{y} $ is the means of the empirical degree distributions: $ \frac{1}{n}\sum\limits_{k}y_k $.
	A different methodology for estimating ERC20 network $ \gamma $ values and their goodness-of-fit can be found in \cite{somin2020network}.

	\subsection{Oscillation Dynamics}

	We consider the ERC20 system as a social physical system and thus use physical models to analyze it.
	We hypothesize that the ERC20 system behaves as a dynamical system approaching its equilibrium state, which can be modeled as a damped harmonics oscillator.
	
	A harmonic oscillator is a system acted upon by a force negatively proportional to its perturbation from its equilibrium state.
	Physical systems that are modeled in this way are springs and swings.
	Systems that also experience a velocity-dependent friction-like force, e.g. air resistance, are modeled by a damped harmonic oscillator.
	The dynamical equation for these models is:
	\begin{equation}
	m\frac{d^2x}{dt^2} = -kx-c\frac{dx}{dt}
	\end{equation}
	
	where $x$ is the perturbation from equilibrium, $m$ is the mass, $k$ is the spring constant and $c$ is the viscous damping coefficient.
	The resonant frequency of the system is defined as $\omega_0=\sqrt{m/k}$ and represents the oscillation of an {\em undamped} system.
	One can define the damping ratio as $\zeta=\frac{c}{2\sqrt{mk}}$ which represents how strong the damping is, compared to the resonant frequency, such that an over-damped system $\zeta>1$ does not oscillate, but exponentially converges to the equilibrium state, whereas an under-damped system $\zeta<1$ oscillates with a modified frequency $\omega_1=\omega_0\sqrt{1-\zeta^2}$ during its exponential convergence. The case of critically damped system $\zeta=1$ is an important one in physics, but does not relate to the analysis presented below.
	
	Given an under-damped oscillator, the dynamics of the system can be described by the following function:
	\begin{equation}
	x(t) = A\cdot e^{-\omega_0\zeta t}\cdot\sin(\omega_0\sqrt{1-\zeta^2}t+\varphi)+x_{\infty}  
	\end{equation}
	Here $\varphi$ is the phase of the oscillation and $x_\infty$ is the equilibrium state. 
	
	In this paper, we will use the under-damped oscillator in order to  model the dynamics of the ERC20 network meta-parameter $\gamma$ and extract the parameters of its dynamics.

	\section{Results}
	\label{sec.results}
	
	In this work we analyze the dynamics of ERC20 tokens' trading over the Ethereum Blockchain. 
	We obtain the ERC20 transactions using the methodology thoroughly explained in \cite{somin2018social}. 
	We have retrieved all ERC20 tokens transactions spreading between February 2016 and June 2018, resulting in $88,985,493 $ token trades, performed by $17,611,649$ unique wallets,  trading $ 51,281 $ token addresses. 
	
	During the examined timespan of $ 2.5 $ years of ERC20 transactions, the economy keeps evolving and changing its dynamics. 
	 Not only does the rising public interest in Blockchain and tokens induce an exponential growth in transactions' volume, but the traded tokens in this economy change as well, as new tokens are established and others lose their impact and decay. 
	 A thorough discussion of the dynamics of the economical properties of the ERC20 economy was conducted in \cite{somin2020network}.
	
	\subsection{Temporal Dynamics: The Oscillating Network Model}
	This apparent volatile nature of ERC20 ecosystem leads us to examine its network characteristics along time. 
	We therefore apply temporal graph analysis to a sliding window of weekly graphs. 
	Namely, we define a weekly transactions graph $ G_t $ based on ERC20 trading activity during a week $ [t-7,\,t) $ as follows: 
	
	\begin{definition}\label{def:week_trans_graph}
		The weekly transactions graph for a given day $t$, $G_t(V_t,E_t)$ is the directed graph constructed from all trading transactions over any ERC20 token, made during the time period $ [t-7,t) $. 
		The set of vertices $V_t$ consists of all wallets trading during that period: 
		\begin{equation}
		V_t:=\left\{v\|\hspace{0.1cm} \textnormal{wallet } v  \textnormal{ bought or sold any token during } [t-7,t) \right\}
		\end{equation}
		and the set of edges $ E_t \subseteq V_t\times{V_t}$  is defined as:
		\begin{equation}
		E_t :=\left \{
		(u,v)\| \hspace{0.1cm}  \textnormal{wallet } u \textnormal{ sold to wallet }  v  \textnormal{ any token during } [t-7,t) \right \}
		\end{equation}
	\end{definition} 
	Over the examined period of 2.5 years, we construct 1000 such weekly transactions graphs, using daily rolling windows, each containing one week of transactional data. 
	As seen in \cite{somin2018network} the incoming and outgoing degree distributions of $ G_t $ clearly adhere to a power-law model.   
	
\begin{figure}[H]
		\centering
		\subfloat {{\includegraphics[height=2.62cm]{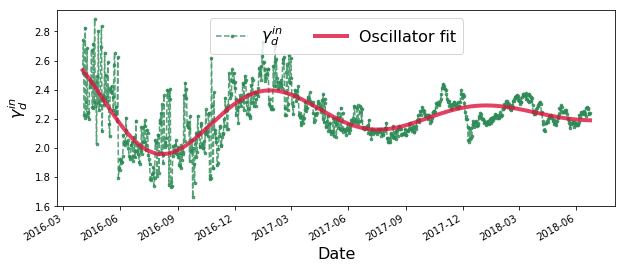} }}%
		\subfloat{{\includegraphics[height=2.62cm]{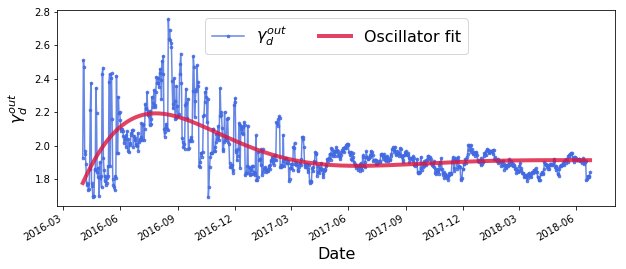} }}%
		\caption{ERC20 transactional network temporal development, in a network related prism, demonstrating the underlying consolidation process the network undergoes. Evolvement of incoming degree distribution gradient, $\gamma_t^{in} $, is depicted in the upper panel and out-degree distribution gradient $ \gamma_t^{out} $ is displayed in the lower panel. Both gradients converge to their \textit{stable states} $ \gamma_{\infty}^{in} $ and $ \gamma_{\infty}^{out} $ correspondingly, following a Harmonic Under-Damped Oscillator model.  }%
		\label{fig:oscillator}
	\end{figure}

	Next, we turn to model the degree distribution over time, as captured by its associated $ \gamma $ values.  
	We postulate that any network of human related transactions, has a characteristic \textit{stable state}, in the form of $ \gamma^{in}_{\infty} $ and $ \gamma^{out}_{\infty} $, to which the network strives to converge:
	\[
	\gamma_t^{in} \underset{t\rightarrow{\infty}}{\longrightarrow}\gamma_{\infty}^{in},\,\,\, \gamma_t^{out} \underset{t\rightarrow{\infty}}{\longrightarrow}\gamma_{\infty}^{out}
	\]
	Empirical observations of both $\gamma_t^{in} $ and $\gamma_t^{out} $ coincide with this hypothesis, as can be seen in Fig. \ref{fig:oscillator}, and can be efficiently modeled as an Harmonic Under-Damped Oscillator, formally $ \forall t $:
	\[
	osc(t) = A\cdot e^{-\omega_0\zeta t}\cdot\sin(\omega_0\sqrt{1-\zeta^2}t+\varphi)+\gamma_{\infty} 
	\]

\begin{figure}[H]
			\hspace{3cm}\subfloat {{\includegraphics[height=4.cm]{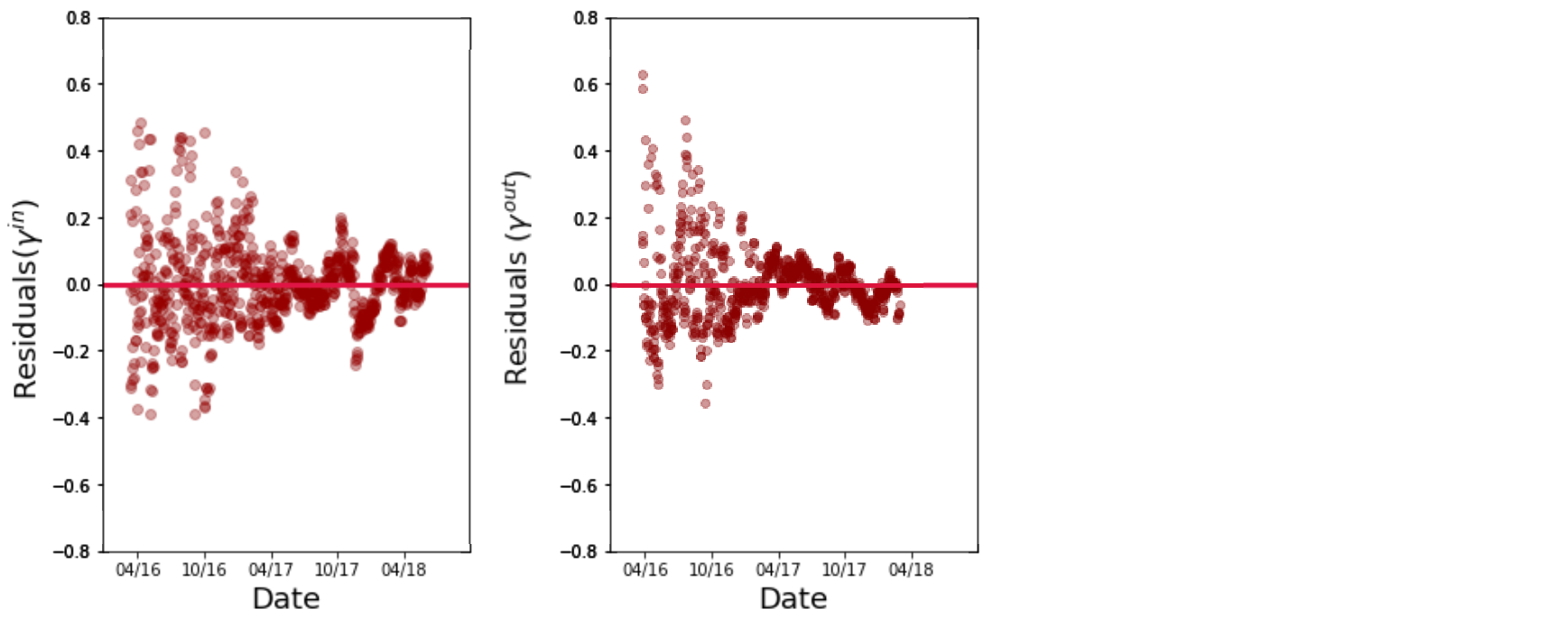} }}%
			\caption{Residuals plots against time for the Oscillator model's fit to $ \gamma^{in} $ and $ \gamma^{out} $. Both demonstrate a symmetrical dispersion around a zero mean.}
			\label{fig:oscillator_residuals}
	\end{figure}

	In order to examine how well the Oscillator model describes and models the dynamics of the degree distribution along time, we explore the residuals from the fit, i.e the deviations of the dependent variable, $ osc_{FT} $, from the fitted oscillator for each day, $ t\in FT $:
		\begin{equation}
		Residual(osc(t)) = \gamma(t) - osc(t) 
		\end{equation}
	Fig. \ref{fig:oscillator_residuals} presents the residuals of the oscillator fit for both $ \gamma^{in} $ and $ \gamma^{out} $. 
	These demonstrate a symmetrical dispersion around a zero mean, proving the validity of modeling the empirical data using an under-damped oscillator. 
	 The residual plots against time further exhibit another interesting phenomena, presenting a decreasing standard deviation of the residuals values along time.
	
	The under-damped oscillator model can be considered as an extension of the regular single-parameter model, which suggested $ \gamma  $ as a constant \textit{stable state}, to a new model governed by five parameters: (\textit{i})  $\lambda = \omega_0\zeta $ representing the exponential decay, (\textit{ii}) $ \omega = \omega_0\sqrt{1-\zeta^2}$ standing for the angular frequency, (\textit{iii}) $ \gamma_{\infty} $ for the stable state to which the system converges, (\textit{iv}) $ A $ representing the maximal amplitude of the oscillation and (\textit{v}) $ \varphi $ for the phase shift. The parameters of fitted oscillators to  $\gamma_t^{in} $  and  $\gamma_t^{out} $, $ osc^{in}(t) $ and $ osc^{out}(t) $ correspondingly, are presented in Table \ref{tab:oscillator_fit} and Table \ref{tab:oscillator_fit_derived_params}.

	\begin{table}[H]   
		\begin{center}
			\caption{\label{tab:oscillator_fit} \small
				Under-Damped Oscillator Models Parameters}
			
			\begin{tabular}{rrrrrr} 
				
				\hline
				\\
				\textbf{Type}  & $ \qquad\qquad \boldsymbol{A} $ &$ \qquad\qquad \boldsymbol{\varphi } $ &$ \qquad\qquad \boldsymbol{\gamma_{\infty} }$ &$ \qquad\qquad \boldsymbol{\frac{2\pi}{\omega_0} }$ \textbf{(days)}& $ \qquad\qquad\boldsymbol{\zeta }$ 		 			     \\
				[0.1cm]
				\hline
				\hline \noalign{\smallskip}
				$\boldsymbol{ osc_{FT}^{out}(t) }$  & -0.77 & 2.96 & 1.91 & 530.2 &0.577\\[0.1cm]
				\hline \noalign{\smallskip}
				$ \boldsymbol{osc_{FT}^{in}(t)}$& 0.39 & 2.23 & 2.23 & 341.1 & 0.152\\[0.1cm]
				\hline
			\end{tabular}
			
		\end{center}
	\end{table}

	\vspace{-1cm} 
	\begin{table}[H]   
		\begin{center}
			\caption{\label{tab:oscillator_fit_derived_params} \small
				Under-Damped Oscillator Models Derived Parameters}
			
			\begin{tabular}{rrrrr} 
				
				\hline
				\\
				\textbf{Type}  & $ \qquad\qquad \boldsymbol{\frac{1}{\lambda}}  $ \textbf{(days)}&$ \qquad\qquad \boldsymbol{\frac{2\pi}{\omega} } $ \textbf{(days)} &$ \qquad\qquad \boldsymbol{k}$ &$ \qquad\qquad \boldsymbol{c }$		 			     \\
				[0.1cm]
				\hline
				\hline \noalign{\smallskip}
				$\boldsymbol{ osc_{FT}^{out}(t) }$  & 146.3 & 649.1 & $1.40\mathrm{e}{-4}$&  $1.37\mathrm{e}{-2}$\\[0.1cm]
				\hline \noalign{\smallskip}
				$ \boldsymbol{osc_{FT}^{in}(t)}$& 356.6 & 345.6 & $3.38\mathrm{e}{-4}$& $5.61\mathrm{e}{-3}$\\[0.1cm]
				\hline
			\end{tabular}
			
		\end{center}
	\end{table}
	
	As part of this novel approach to modeling the network's consolidation process, one should further note that the amplitude of the under-damped oscillator is governed by:  
	\begin{equation}
	A\cdot e^{-\omega_0\zeta t}
	\end{equation}
	The latter enables establishing the time $ t_1 $ at which the network has reached a \textit{stabilization of }$ x\% $, formally:
	
	\begin{definition}\label{def.stabilization_time}
		Let $ G_t(V_t, E_t) $ be the directed graph based on all transactions made during $ [t-7,\,t) $, trading any of the ERC20 tokens, for a given $ t\in FT $.
		Let $ \gamma_t $ denote the power of the associated degree distribution of $ G_t $, whose dynamics modeled by an oscillator $ \gamma_{fit} $. We define the '$ x \% $ stabilization time of the network' w.r.t $ \gamma $ to be the time $ t_1 $ when the amplitude of $ \gamma_{fit} $ reaches at most $ x \% $ of the initial amplitude, observed as time $ t_0 $:
		\begin{equation}
		\begin{aligned}
		x=e^{\omega_0 \zeta(t_0-t_1)} \,& \,\Longrightarrow \,t_1 =t_0-\frac{\ln(x)}{\omega_0\zeta}
		\end{aligned}
		\end{equation}
	\end{definition}	
	This, in turn, enables us to establish the time required for the network to reach stabilization, in both aspects of $\gamma_t^{in} $ and $\gamma_t^{out} $.
	For instance, using the fitted parameters of the under-damped oscillator depicted in Table \ref{tab:oscillator_fit}, one can verify that a 70\% stabilization occurs after 430 days for $ \gamma_t^{in} $:
	\begin{align}
	t_1^{\gamma^{in}} &= -\frac{\ln(0.3)}{0.018\cdot0.152}\nonumber=429.3 \nonumber
	\end{align}
	Using $ x=0.3,\,\omega_0=0.018,\,\zeta=0.152,\,t_0=0 $.
	$ \gamma_d^{out} $ presents the same stabilization after merely 177 days:
	
	\begin{align}
	t_1^{\gamma^{out}} &= -\frac{\ln(0.3)}{0.011\cdot0.577}\nonumber=176.1 \nonumber
	\end{align}
	where $ x=0.3,\,\omega_0=0.011,\,\zeta=0.577,\,t_0=0 $.

\subsection{The Oscillating Network Model:  Predictive Ability}

	Once the modeling of the ERC20 network dynamics by an under-damped oscillator is established, it can be also used for predictive purposes. With this objective in mind, we fit partial $ \gamma $ observations to an oscillator model, considering data restricted by date, in order to predict future $ \gamma $ dynamics, formally defined as	:
	\begin{definition}
		Let $ T_0 $ stand for the minimal date available in our dataset, April 1st, 2016. 
		Given any time-stamp $ T_i>T_0  $, we define $osc_{T_i} $ to be the partial oscillator model, representing  the under-damped oscillator model fitted to $ \gamma $ values between $ T_0 $ and $ T_i $. The parameters characterizing $osc_{T_i} $ are denoted by  $ A(T_i), \,\varphi(T_i),\,\gamma_\infty(T_i),\,\omega_0(T_i)$ and $\zeta(T_i) $.
	\end{definition}

	In this constellation, $ T_i $ is incremented on a daily basis, starting from April 28th, 2016, resulting in a set of \textit{partial oscillator models}, each fitted to $ \gamma  $ values occurring between $ [T_0,\,T_i) $: 
	\[
	\bigcup_{T_i}   osc_{T_i} 
	\]
	The initial parameters values and their corresponding upper and lower bounds, as supplied to the oscillator model in the fitting process, are depicted in Table \ref{tab:partial_oscillator_in_init_params}.
	
	\begin{table}[H]   
		\begin{center}
			\caption{\label{tab:partial_oscillator_in_init_params} \small
				Partial Oscillator Models Fit Parameters}
			
			\begin{tabular}{rrrrrr} 
				
				\hline
				\\
				\textbf{Type}  & $ \qquad \boldsymbol{A} $ &$ \qquad\qquad \boldsymbol{\varphi} $ &$ \qquad\qquad \boldsymbol{\gamma_{\infty} }$ &$ \qquad \boldsymbol{\frac{2\pi}{\omega_0} }$ \textbf{(days)}& $\qquad \qquad\qquad\boldsymbol{\zeta }$ 		 			     \\
				[0.1cm]
				\hline
				\hline \noalign{\smallskip}
				$\boldsymbol{osc_{T_i}^{in} }$ \textbf{ initial values}  & 0.5 & $\pi / 2$& 2.1 & 365 &0.25\\[0.1cm]
				\hline \noalign{\smallskip}
				$\boldsymbol{osc_{T_i}^{out} }$ \textbf{ initial values}  & -0.5 & $\pi / 2$& 1.9 & 365 &0.5\\[0.1cm]
				\hline \noalign{\smallskip}
				\textbf{Bounds}& [-1, 1] & [0, $\pi$] & [1, 3] & [1, 700] &   [0.001, 0.999]\\[0.1cm]
				\hline
			\end{tabular}
			
		\end{center}
	\end{table}

	We start by examining the stabilization process of the partial models' parameters. In order to smoothen their dynamics, we calculate a $ 90 $-days rolling mean over each parameter, retrieved from $ 90 $ consecutive partial oscillator fits. 
	Formally:
	\begin{definition}
		 given a time-stamp $T_i$ and given an oscillator's property  $ P_{osc} $ such that:
		\begin{equation}
		P_{osc} \in \{A, \,\varphi,\,\gamma_\infty,\,\omega_0,\, \zeta  \}
		\end{equation}
		We define the mean and standard deviation of $ P_{osc} $ w.r.t to $ T_i $ as follows: 
		\begin{equation}
		\begin{aligned}
		\mathrm{mean}_{T_i}(P_{osc}) \equiv \underset{t \in [T_i-90,T_i)}{\mathrm{mean}} (P_{osc}(t))\\
		\mathrm{STD}_{T_i}(P_{osc}) \equiv \underset{t \in [T_i-90,T_i)}{\mathrm{STD}} (P_{osc}(t))
		\end{aligned}
		\end{equation}
	\end{definition}
	
\begin{figure}[H]
		\centering
		\subfloat {{\includegraphics[width=6.cm]{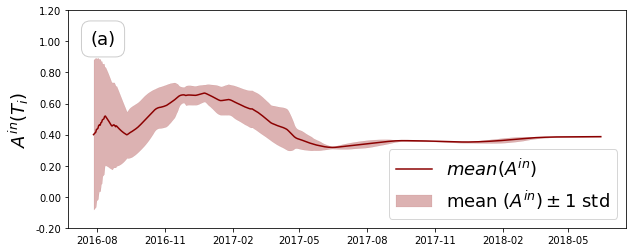} }}
		\hspace*{\fill}
		\subfloat {{\includegraphics[width=6.cm]{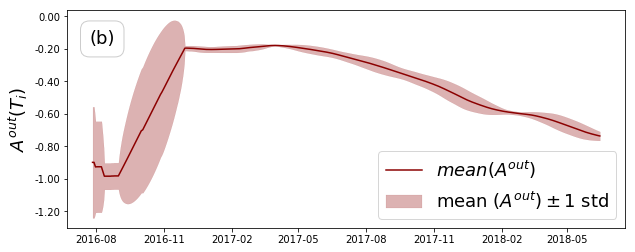} }}
		\hspace{5mm}
		\subfloat{{\includegraphics[width=6.cm]{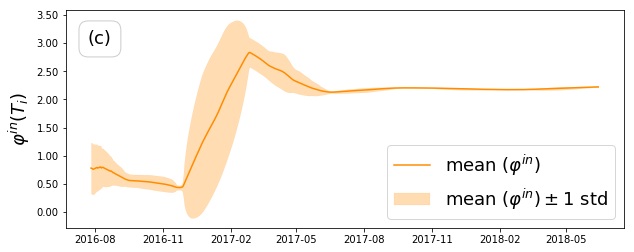} }}
		\hspace*{\fill}
		\subfloat {{\includegraphics[width=6.cm]{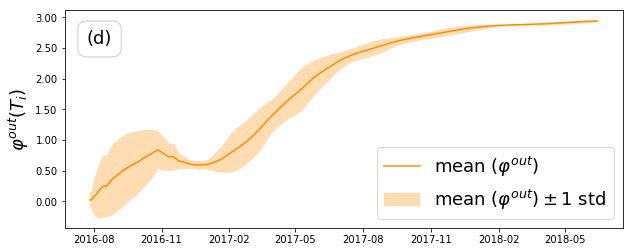} }}
		\hspace{5mm}
		\subfloat{{\includegraphics[width=6.cm]{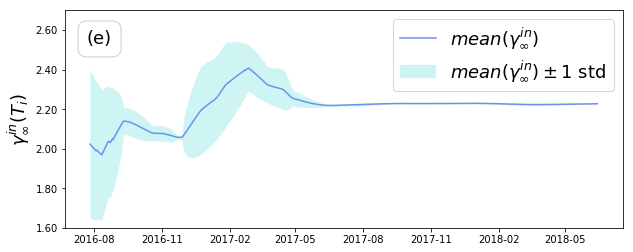} }}
		\hspace*{\fill}
		\subfloat{{\includegraphics[width=6.cm]{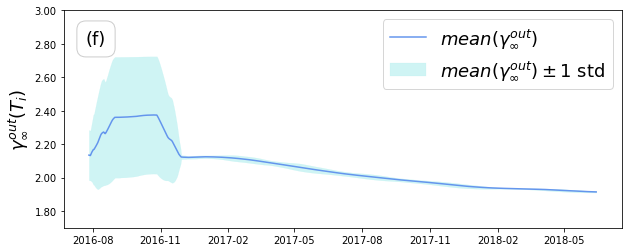} }}%
		\hspace{5mm}
		\subfloat {{\includegraphics[width=6.cm]{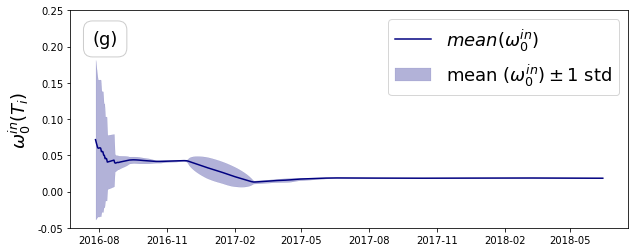} }}
		\hspace*{\fill}
		\subfloat {{\includegraphics[width=6.cm]{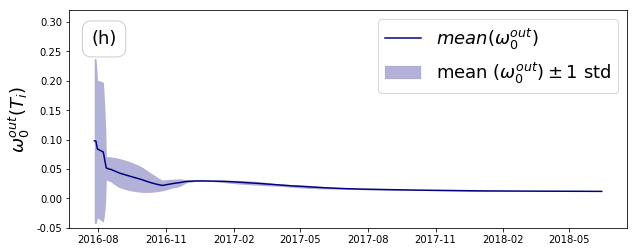} }}
		\hspace{5mm}
		\subfloat {{\includegraphics[width=6.cm]{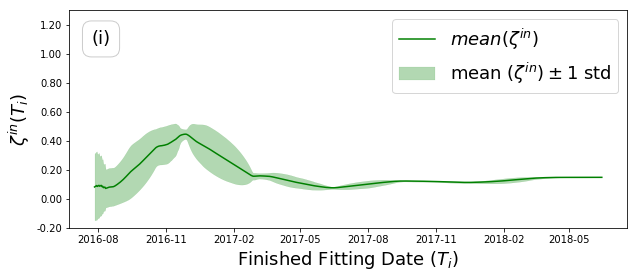} }}
		\hspace*{\fill}
		\subfloat {{\includegraphics[width=6.cm]{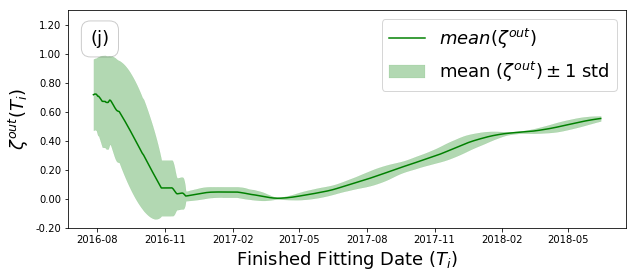} }}
		\caption{Stabilization process of $ osc_{T_i}^{in} $ and $ osc_{T_i}^{in} $ parameters, depicting the mean and standard deviation of $ A $, $ \varPhi $, $ \gamma_\infty $, $ \omega_0 $ and $ \zeta $ for the partial oscillator models fitted on $ \gamma^{in} $ (left panels) and on $ \gamma^{out} $ (right panels). 	All  $ osc_{T_i}^{in} $ parameters present a powerful converging pattern, occurring at June 2017.  $ osc_{T_i}^{out} $ parameters, though presenting a decreasing STD along time, indicating the agreement level of consecutive partial models, do not manifest the same converging ability, as the parameters' values keep changing along time.}
		\label{fig:fitting_osc_params}
	\end{figure}

	Fig. \ref{fig:fitting_osc_params} depicts the stabilization process all parameters undergo along time, as $ T_i $ advances, both for $ \gamma^{in} $ and $ \gamma^{out} $, presenting the $\mathrm{mean}_{T_i}(P_{osc}) $ and   $\mathrm{STD}_{T_i}(P_{osc})  $ along time.
	
	The great difference between $ \gamma^{in} $ and $ \gamma^{out} $ is quite evident in this analysis as well, and is manifested through the parameters' convergence properties.
	Each parameter $ P_{osc} $ of $ osc_{T_i}^{in}  $, the \textit{partial oscillator models} for $ \gamma^{in} $, presents not only a decreasing $ \mathrm{STD}_{T_i}(P_{osc})  $ as $ T_i $ progresses, manifesting the early consensus established by consecutive $ osc_{T_i}^{in}  $-s, but also a clear-cut stabilization for each of the parameters' mean value, formulating at June 2017. This stabilization, occurring approximately a year prior to the end of our data, strongly implies the potentially extraordinary predictive abilities of the model, applied to $ \gamma^{in} $.    
	The parameters of the \textit{partial oscillator models} for $ \gamma^{out} $, although presenting a decreasing standard deviation along time, do not display the same converging tendency as $ T_i $ progresses, manifested by a constant change in the parameters' mean value along time.    
	
	This analysis leads us to examine the predictive abilities of $ osc_{T_i} $, and the amount of data required for fitting the oscillator's parameters, in order to establish a stable and accurate prediction. For this purpose, we select $ 5 $ different \textit{inspection dates}, referred as $e_d$: 
	\begin{enumerate}
		\item [-]September 28th, 2016 
		\item[-] December 27th, 2016
		\item [-]March 27th, 2017
		\item [-]June 25th, 2017
		\item [-]September 23rd, 2017
	\end{enumerate}      
	and analyze the predictive ability of $ osc_{T_i} $' as to $ \gamma $ dynamics over the $ [e_d,\textnormal {June 2018}) $ timespan.

	In order to establish the confidence levels for each $ e_d $ associated prediction, we analyze the performance of $ 90 $ \textit{partial oscillator models} for each $ e_d $, forming a set, we'd refer to as $ O_{e_d} $:
	\begin{equation}
	O_{e_d}  = \bigcup_{T_i\in[e_d-90,e_d)}   osc_{T_i}
	\end{equation}
	The mean and standard deviation of $ O_{e_d} $'s prediction of $ \gamma $ for a given time $ t\in FT $ are defined as:
	\begin{equation}\label{eq:O_e_d}
	\begin{aligned}
	\mathrm{mean}(O_{e_d})(t) \equiv \underset{T_i\in[e_d-90,e_d)}{\mathrm{mean}} (osc_{T_i}(t))\\
	\mathrm{STD}(O_{e_d})(t) \equiv \underset{T_i\in[e_d-90,e_d)}{\mathrm{STD}} (osc_{T_i}(t))
	\end{aligned}
	\end{equation}
	
\begin{figure}[H]
		\centering
		\subfloat {{\includegraphics[width=6.05cm]{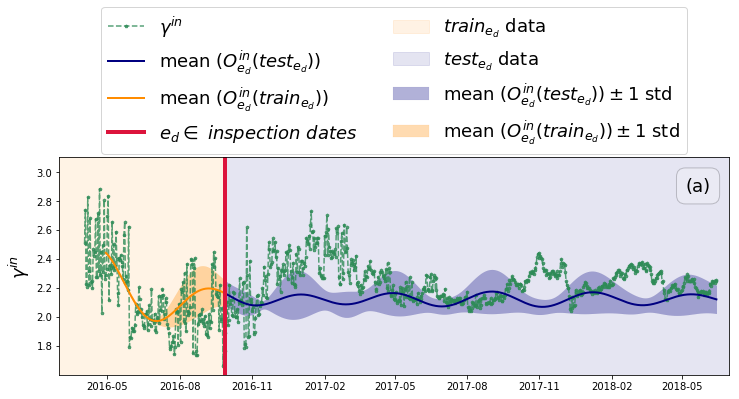} }}
		\hspace{-2.2mm}
		\subfloat {{\includegraphics[width=6.05cm]{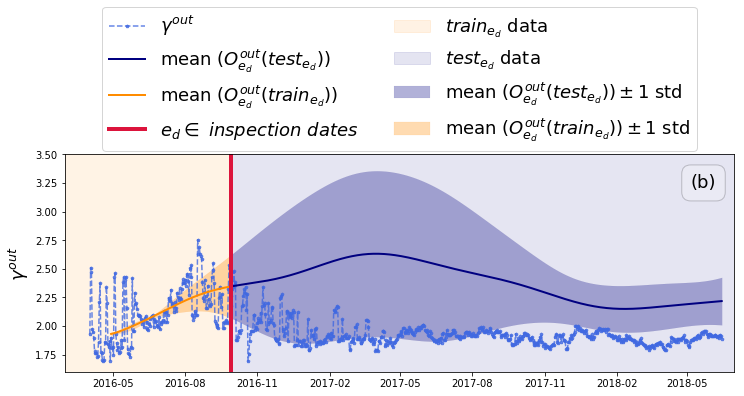} }}
		\hspace{1mm}
		\subfloat{{\includegraphics[width=6.05cm]{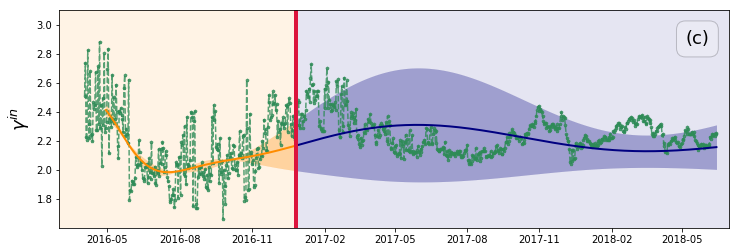} }}
		\hspace{-2.2mm}
		\subfloat {{\includegraphics[width=6.05cm]{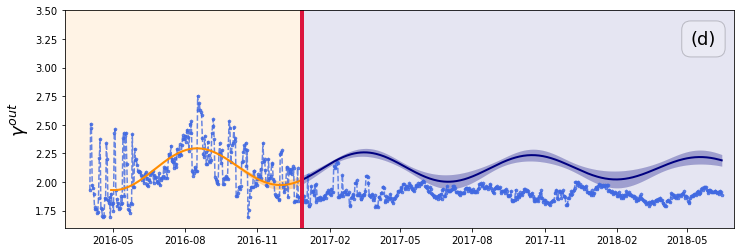} }}
		\hspace{2mm}
		\subfloat{{\includegraphics[width=6.05cm]{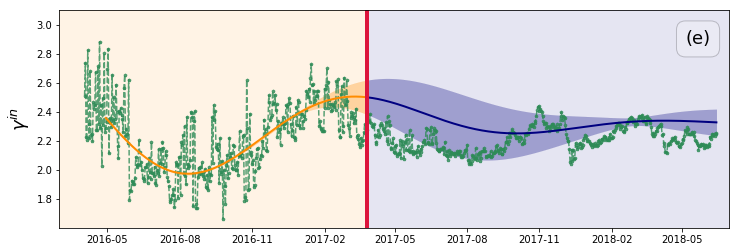} }}
		\hspace{-2.2mm}
		\subfloat{{\includegraphics[width=6.05cm]{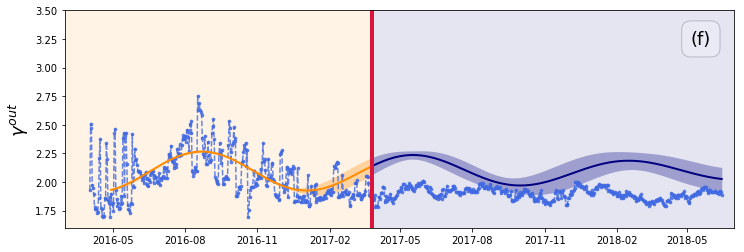} }}%
		\hspace{2mm}
		\subfloat {{\includegraphics[width=6.05cm]{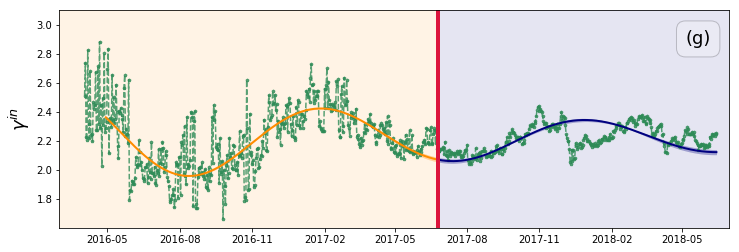} }}
		\hspace{-2.2mm}
		\subfloat {{\includegraphics[width=6.05cm]{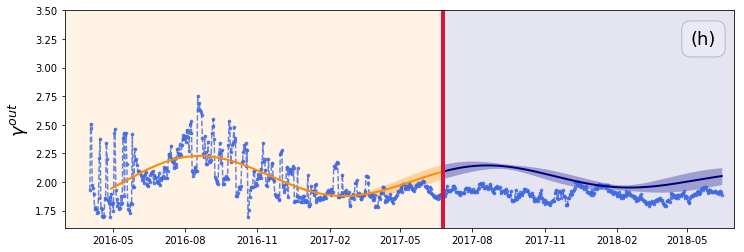} }}
		\hspace{2mm}
		\subfloat {{\includegraphics[width=6.05cm]{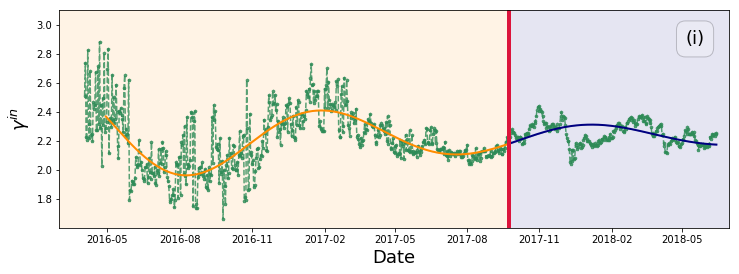} }}
		\hspace{-2.2mm}
		\subfloat {{\includegraphics[width=6.05cm]{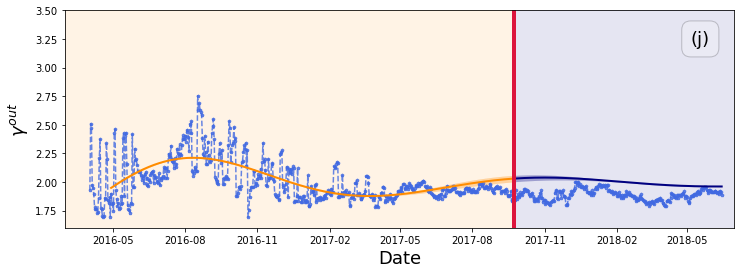} }}
		\caption{
			Analyzing the abilities of the partial oscillator model to predict $ \gamma^{in} $ and $ \gamma^{out} $ dynamics over $ [e_d, $ June 2018), along 5 different inspection dates $ e_d $. Presenting mean and standard deviation of $ O_{e_d} $, the set of 90 consecutive partial oscillator models, fitted until $ e_d $ (see Eq. \ref{eq:O_e_d}). 
			Left panels depicted prediction of $ \gamma^{in} $  dynamics, and right panels present $ \gamma^{out} $ dynamics predictions by $\mathrm{mean}(O_{e_d}^{in})  $ and  $\mathrm{mean}(O_{e_d}^{out})  $ respectively.}
		\label{fig:predict_gamma}
	\end{figure}

	Fig. \ref{fig:predict_gamma} depicts the predictions made for both $ \gamma^{in} $ and $ \gamma^{out} $, presenting $ \mathrm{mean}(O_{e_d}) $ and $ \mathrm{STD}(O_{e_d}) $ along $ [e_d,\textnormal {June 2018}) $, for each of the $ 5  $ \textit{inspection dates}.
	The prediction analysis coincides with our parameters stabilization analysis, as predictions for $ \gamma^{in} $ stabilize as $ e_d $ advances, until finally presenting high reliability, for predicting a whole year of data, starting from $ e_d =$ June 25, 2017. 
	We further observe that confidence levels for the predictions increase with $ e_d $ for both $ \gamma^{in} $ and $ \gamma^{out} $, manifested by the decreasing standard deviation as $ e_d $ progresses. 
	
	We note however, as was also implied by $ \gamma^{out} $ parameters stabilization process, that the predictive ability for $ \gamma^{out} $ when estimated at $ e_d =$ June 25, 2017 isn't as strong, manifested by the evident over-estimation produced by  $ \mathrm{mean}(O_{e_d}) $ while predicting $ \gamma^{out} $ values during $ [\textnormal {June 2017},\,\,\textnormal {June 2018}) $. This apparent bias in the prediction of $ \gamma^{out}$ values manifests the lack of oscillations in the actual observed $ \gamma^{out} $ values, as well as an overestimation of $ \gamma_{\infty} $ by $ \mathrm{mean}(O_{e_d}) $. This may suggest that there are other \textit{forces} influencing the dynamics of $ \gamma^{out} $, rather than just \textit{'spring and friction'}-like forces.
	
	We further wish to analyze the \textit{'goodness-of-fit'} of $ osc_{T_i} $ for any given $ T_i $, for the entire $ [T_i,\textnormal {June 2018}) $ timespan. 
	We therefore calculate the \textit{Root Mean Squared Error} of each $ osc_{T_i} $, namely:
	\begin{equation}
	\mathrm{RMSE}(osc_{T_i}) = \sqrt{\frac{1}{n_{T_i}}\underset{t\in [T_i, \textnormal {June 2018})}{\sum}(\gamma(t) - osc_{T_i}(t))^2}
	\end{equation}
	where $ n_{T_i} $ is the length of $ [T_i, $ June 2018$)$ period, measured in days. 
	
	In order to smoothen the RMSE signal, we calculate a 90-days rolling mean over the RMSE of the \textit{partial oscillator model}:
	\begin{equation}
	\begin{aligned}
	\mathrm{mean}_{T_i}(\textnormal{RMSE}(osc) ) \equiv \underset{t \in [T_i-90,T_i)}{\mathrm{mean}} (\textnormal{RMSE}(osc_{t}) )\\
	\mathrm{STD}_{T_i}(\textnormal{RMSE}(osc) ) \equiv \underset{t \in [T_i-90,T_i)}{\mathrm{STD}} (\textnormal{RMSE}(osc_{t}) )\\
	\end{aligned}
	\end{equation}
	
\begin{figure}[H]
		\centering
		\subfloat {{\includegraphics[width=10.cm]{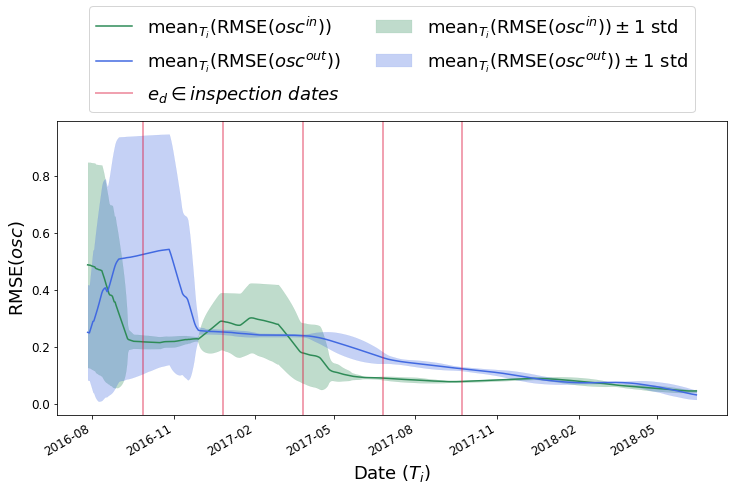} }}
		\caption{Average and standard deviation of Root Mean Squared Error of both $ osc_{T_i}^{out} $ and $ osc_{T_i}^{in} $, over 90 consecutive partial oscillator models, $ osc_{T_i} $, manifesting the ability of the under-damped oscillator to predict $ \gamma $ dynamics. This analysis presents both the improving accuracy of models as time advances, and their growing stability, manifested by a decreasing standard deviation.}
		\label{fig:RMSE}
	\end{figure}

	Fig. \ref{fig:RMSE} depicts the Root Mean Squared Error of both $ osc_{T_i}^{in} $ and $ osc_{T_i}^{out} $, presenting  both $ \mathrm{mean}_{T_i}(\textnormal{RMSE}(osc) )  $ and $ \mathrm{STD}_{T_i}(\textnormal{RMSE}(osc) )  $ for predicting $ \gamma^{in} $ and $ \gamma^{out} $ along time. 
	This analysis assists in validating that in average, $ osc_{T_i}^{in} $ has lower error values compared to $ osc_{T_i}^{out} $ along most of the examined timespan and specifically over the predictions of the last year of data, starting from  $ e_d =$ June 25, 2017. 
	
	We note however that both $ osc_{T_i}^{in} $ and $ osc_{T_i}^{out} $ converge to similar RMSE values, starting from December 2017, yielding 7 months of similar and low error predictions for both $ \gamma^{in} $ and $ \gamma^{out} $ dynamics. The decreasing RMSE, and its standard deviation enhance even further the predictive abilities of the under-damped oscillator as a model of $ \gamma $ dynamics, concluding that $ osc^{in} $ is a better predictor compared to $ osc^{out} $ , as both its error rate and its confidence levels, present significant decrease earlier in time, enabling a reliable prediction of an entire year of data.

	\section{Discussion} \label{sec.discussion}
	In this contribution, we aimed to go beyond a static view of the ERC20 ecosystem, and explore its dynamics along time.
	Specifically, we have chosen to focus on the dynamics analysis of the most basic and the highly investigated network characteristic of them all, studying the development of the network's degree distribution, manifested by its associated power, $\gamma$, throughout time. 
	Inspired by macro-scale market dynamics \cite{frisch1933propagation,goodwin1993economy}, which demonstrated an oscillatory stabilization process, we analyzed the exponents of in and out-degree distribution ($ \gamma^{in} $ and $ \gamma^{out} $) and studied to which extent their dynamics can be modeled by an under-damped harmonic oscillator.
	
	The goodness of fit of the oscillator model  to $ \gamma^{in} $ was tested by analyzing the residuals plots, verifying they were centered around zero (see Fig. \ref{fig:oscillator_residuals}.
	 Moreover, Fig.~\ref{fig:fitting_osc_params} (left panels) demonstrates how the fitted parameters of the oscillator model fitted to $ \gamma^{in} $ stabilize together and early, prior to the last observable oscillation in actual $ \gamma^{in} $ values.
	We further demonstrated the powerful predictive ability of the oscillator model for $ \gamma^{in} $. 
	Predicting future values based on fitting to an under-damped oscillator, Fig.~\ref{fig:predict_gamma}(left-panels) shows that damped oscillations during the last year of data can be accurately and reliably predicted.

	\bibliography{prediction_gamma}
	\bibliographystyle{ieeetr}
\end{document}